\documentclass[a4paper,12pt]{spieman}
\usepackage{amsmath,amsfonts,amssymb}
\usepackage{graphicx}
\usepackage{setspace}
\usepackage{tocloft}
\usepackage{textcomp}
\usepackage{tocloft}
\usepackage[flushleft]{threeparttable}
\usepackage{graphicx}
\usepackage{lipsum}
\usepackage{float}
\usepackage{amsmath}
\usepackage{latexsym}
\usepackage{booktabs,etoolbox}
\usepackage[utf8x]{inputenc}

\newcommand*\aap{A\&A}
\newcommand*\aapr{A\&A~Rev.}
\newcommand*\aaps{A\&AS}

\newcommand*\aj{AJ}

\newcommand*\apj{ApJ}
\newcommand*\apjl{ApJ}

\newcommand*\araa{ARA\&A}

\newcommand*\mnras{MNRAS}

\newcommand*\nat{Nature}

\newcommand*\pasp{PASP}

\newcommand*\ssr{Space~Sci.~Rev.}

\title{Colorado Ultraviolet Transit Experiment Data Simulator}
\author[a]{Aickara Gopinathan Sreejith}
\author[a]{Luca Fossati}
\author[b]{Brian T. Fleming}
\author[b]{Kevin France}
\author[c]{Tommi Koskinen}
\author[b]{Arika Egan}
\author[a]{Hannah T. R\"udisser}
\author[a]{Manfred Steller}

\affil[a]{Space Research Institute, Austrian Academy of Sciences, Schmiedlstrasse 6, 8042 Graz, Austria}
\affil[b]{Laboratory for Atmospheric and Space Physics, University of Colorado, UCB 600, Boulder, CO, 80309, USA}
\affil[c]{Lunar and Planetary Laboratory, University of Arizona, 1629 East University Boulevard, Tucson, AZ 85721-0092, USA}

\cftpagenumbersoff{figure}
\cftpagenumbersoff{table} 
\begin{document}           
%\linenumbers
%\authorrunning{A.G. Sreejith et al.}
%\titlerunning{CUTE data simulator}
\maketitle

\begin{abstract}
The Colorado Ultraviolet Transit Experiment (CUTE) is a 6U NASA CubeSat carrying on-board a low-resolution (R$\sim$ 2000--3000), near-ultraviolet (2500--3300\,\AA) spectrograph. It has a rectangular primary Cassegrain telescope to maximize the collecting area. CUTE, which is planned for launch in Spring 2020, is designed to monitor transiting extra-solar planets orbiting bright, nearby stars aiming at improving our understanding of planet atmospheric escape and star-planet interaction processes. We present here the CUTE data simulator, which we complemented with a basic data reduction pipeline. This pipeline will be then updated once the final CUTE data reduction pipeline is developed. We show here the application of the simulator to the HD209458 system and a first estimate of the precision on the measurement of the transit depth as a function of temperature and magnitude of the host star. We also present estimates of the effect of spacecraft jitter on the final spectral resolution. The simulator has been developed considering also scalability and adaptability to other missions carrying on-board a long-slit spectrograph. The data simulator will be used to inform the CUTE target selection, choose the spacecraft and instrument settings for each observation, and construct synthetic CUTE wavelength-dependent transit light curves on which to develop the CUTE data reduction pipeline.
\end{abstract}


\begin   {spacing}{1.2}   % use double spacing for rest of manuscript

\section{Introduction}
%-------------------------------------
Thanks to technical and instrumental advances, the field of extra-solar planet (hereafter exoplanet) research is steadily moving from a detection to a characterisation phase, focusing particularly on structure and composition of planetary atmospheres. Atmospheric characterisation is possible almost exclusively for transiting exoplanets, thus those for which the orbital geometry is such that planets pass in front of and behind the host star as seen from Earth. During a primary transit, that is when the planet is located between the host star and the Earth, part of the stellar light passes through the planetary atmosphere, which leaves its fingerprints in the stellar spectrum. The difference between the stellar spectra obtained in- and out-of-transit gives then the planetary transmission spectrum e.g., Ref.~\citenum{brown2001}, which can be compared to synthetic transmission spectra to estimate for example the planetary atmospheric structure and chemical composition.

Transmission spectroscopy is currently the main tool employed to characterise planetary atmospheres and it has been carried out from X-ray to radio wavelengths leading to a number of important results and discoveries (see e.g., Refs.~\citenum{fossati2015,crossfield2015}, for reviews). Among those, the most relevant for this work is the detection of significant planet atmospheric escape, that is the loss of atmospheric gas to space\cite{vidal2003,fossati2010}. Since then, it has become clear that atmospheric escape plays a key role in planetary evolution, including solar system planets, and that it is one of the key phenomena shaping the currently observed short-period exoplanet population\cite{johnstone2015,owen2017,haswell2017,jin2018,lammer2018}.

Because of the low optical depth of the escaping gas in the planetary upper atmosphere, escape can be best studied at ultraviolet (UV) wavelengths. The exoplanet UV transit observations conducted so far led to the detection of a large variety of phenomena, such as an excess of absorption during transit over what is observed at optical wavelengths in both narrow and wide wavelength ranges, and transit light curve asymmetries and variability (e.g., Refs.~\citenum{vidal2003,fossati2010,linsky2010,lecavelier2012,haswell2012,benjaffel2013,vidal,kulow2014,ehrenreich2015,ballester2015}). At present, the theories explaining them exceed the number of relevant transit observations. There is a whole wealth of phenomena, also variable in time, that requires a large observational effort to be understood, effort that cannot be undertaken by the Hubble Space Telescope (HST) alone, which is our almost only UV ``eye''. As a matter of fact, the thorough observational program necessary to begin systematically constraining the theories of atmospheric escape would require more than a thousand HST orbits, which cannot be scheduled on a share-used facility.

However, owing to the large size of escaping atmospheres and to the short orbital periods of close-in planets, those most subject to mass loss, the physics of atmospheric escape can be studied with a dedicated small instrument operating at near-UV (NUV) wavelengths (250--320\,nm). Most of the previous observations looking at atmospheric escape concentrated on the stellar emission lines, such as H{\sc i} Ly$\alpha$ (1216\,\AA) and C{\sc ii} (1334\,\AA), as the background light source (see e.g., Ref.~\citenum{fossati2015}, for a review). While these observations are made against faint chromospheric emission lines of the host star in the far-UV (FUV), the NUV spectral range presents a large number of strong resonance lines of abundant metals (e.g., Mg{\sc i}, Mg{\sc ii}, Fe{\sc ii}) observed in absorption against the NUV photospheric continuum, thus having a flux at least ten to hundred times that typical of Ly$\alpha$. 

\begin{figure}[h!]
\begin{center}
\includegraphics[width=\textwidth]{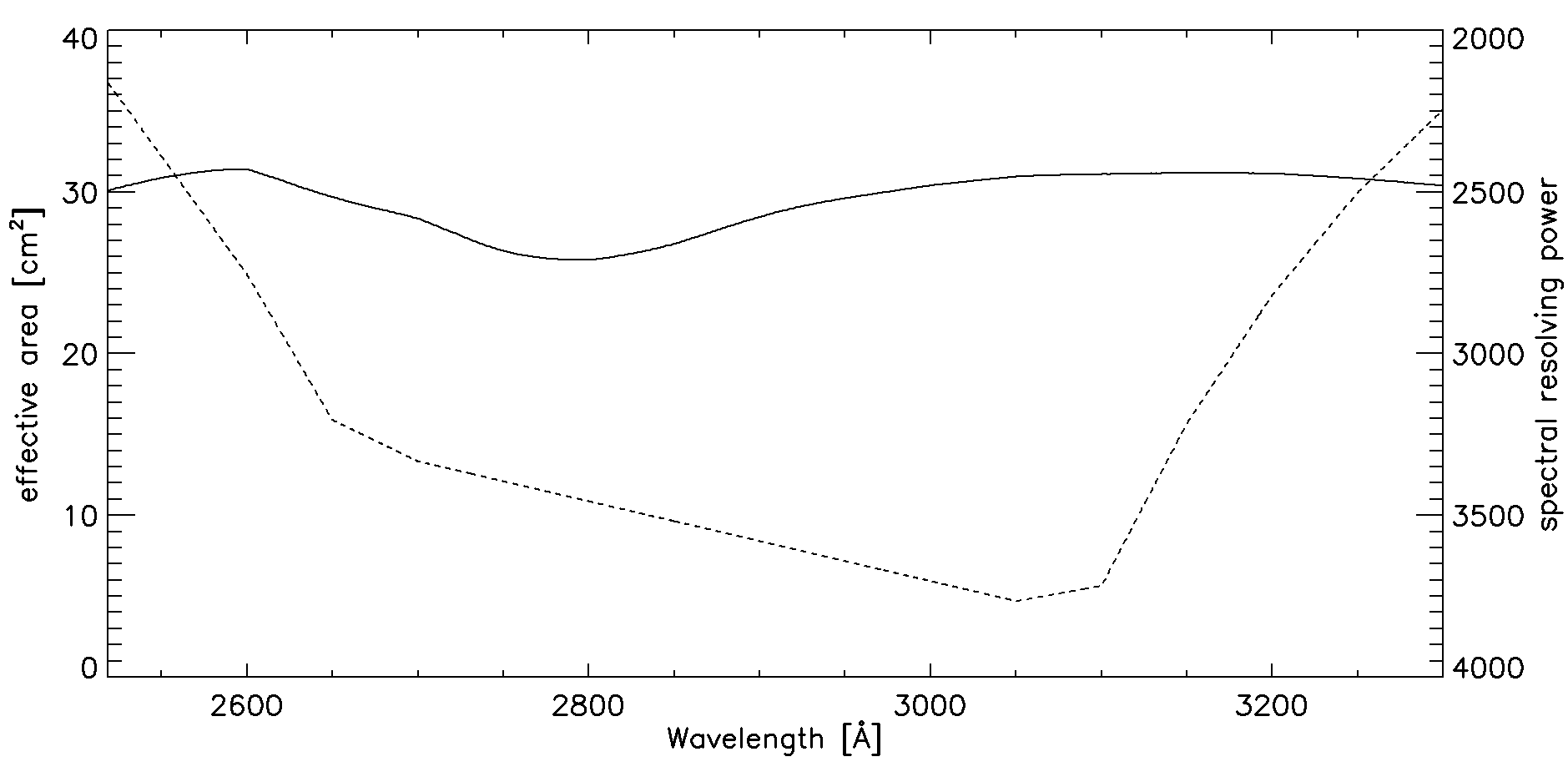}
\caption{CUTE effective area (solid line) and CUTE spectral resolving power at the center of the slit (dashed line).}
\label{fig:fig0}
\end{center}
\end{figure}

The Colorado Ultraviolet Transit Experiment (CUTE) is a 6U CubeSat specifically designed to provide the spectroscopic observations needed to further understand atmospheric escape. In short, the CUTE instrument comprises of a rectangular Cassegrain telescope feeding light into a low-resolution (comparable to that of HST COS G230L) spectrograph with ion etched holographic grating operating from 2500\,\AA\ to 3300\,\AA. The CUTE science instrument is incorporated into a 6U form factor spacecraft. A detailed description of CUTE design, hardware, and operations is provided in Refs.~\citenum{fleming,egan}. We show the CUTE effective area and spectral resolution in Fig:~\ref{fig:fig0}.

%-------------------------------------
%-------------------------------------

We present here the CUTE data simulator, dubbed ACUTEDIRNDL, that has been designed to realistically identify the future capabilities of CUTE, thus, for example, guide target selection. This is a particularly important task given the large number of planets orbiting bright stars that are expected to be found by the many planet-finding facilities currently at work, such as TESS\cite{}, KELT\cite{}, HAT-Net\cite{}, and Mascara\cite{}. 

The basic idea of the simulator was presented in Ref.~\citenum{sreejith}. This paper describes the simulator's architecture and operation (Sect.~\ref{sec:architecture}). We also present here the results of simulations on a number of systems and the effect of spacecraft jitter on the final CUTE data products (Sect.~\ref{sec:results}). In Sect.~\ref{sec:conclusions}, we present the summary of this work and describe the future developments of the simulator.
\section{Simulator Architecture and Operation}\label{sec:architecture}
%-------------------------------------
The simulator\footnote{The simulator is available at: {\tt https://github.com/agsreejith/ACUTEDIRNDL}.} is a set of IDL routines generating images that reproduce spectral time series of stars taking into account the wavelength-dependent planetary absorption during transit and instrumental effects. The simulator recreates the effects that CCD's (e.g., size, pixel scale, and cosmetics), readout electronics, optical elements (telescope and spectrograph), planetary absorption during transit, spacecraft orientation and jitter, and systematic noise sources have on the data. This allows the user to best foresee the data quality and the magnitude of different sources of noise (both white and red). The simulator is fed by a wide range of input parameters providing high flexibility. It follows that the simulator, which is originally designed for CUTE, can be easily adapted to work for any other mission carrying on-board a long-slit spectrograph and a charge transfer device as detector.
\subsection{Input parameters}
The simulator requires a set of input parameters, which are given through the input parameter file.
\begin{itemize}
  \item Stellar parameters: stellar temperature, radius, $V$-band magnitude, and parallax, stellar flux at Earth (optional).
  \item Transit parameters: impact parameter, planetary orbital period, planetary radius, orbital inclination, orbital semi-major axis, mid-transit time, wavelength-dependent shape of the transit light curve (optional). 
  \item Target and exposure parameters: RA and DEC (in degrees), exposure time, time corresponding to the beginning of the first observation (optional).
  \item Instrument parameters: position of the star with respect to the slit center, spectral resolution, instrument effective area, shape of the spectral footprint in the cross-dispersion direction. 
  \item Background parameters: background parameters (STIS or zodiacal), scattered-light parameters (optional), background stars (optional).
  \item CCD parameters: physical size (in pixels), readout noise, dark level, bias level, readout time, CCD gain, plate scale, linearity (optional)
  \item Spacecraft parameters: jitter parameters, orientation, observing gaps (optional), instrument systematics (optional).
\end{itemize}
We provide below a detailed description of each of these parameters.

\noindent{\bf Stellar parameters.} The simulator requires the stellar flux at Earth in erg\,cm$^{-2}$\,s$^{-1}$\,\AA$^{-1}$ as a function of wavelength in Angstrom. This can be provided either directly in a file (e.g., spectrum obtained from another instrument in ascii format) or by providing the stellar temperature (in Kelvin), radius (in solar radii), Johnson apparent $V$-band magnitude, and parallax (in milliarcseconds). 

Based on the input stellar temperature, the simulator extracts the corresponding stellar fluxes from a library of models that we generated employing the {\it LLmodels} stellar atmosphere code\cite{shulyak}, which computed photospheric fluxes of stars assuming local thermodynamical equilibrium (LTE). The library covers stars raging between 3500\,K and 12000\,K, in steps of 100\,K below 6800\,K and in steps of 200\,K above it. Each synthetic spectral energy distribution covers wavelengths between 100\,\AA\ and 30\,$\mu$m with an average wavelength sampling of 1\,\AA, which increases to 0.005\,\AA\ between 1500 and 9000\,\AA. The output stellar fluxes, which include the continuum, are in units of erg\,cm$^{-2}$\,s$^{-1}$\,Hz$^{-1}$. To account for chromospheric emission at the core of the Mg{\sc ii}\,h\&k lines in late-type stars (i.e., cooler than about 6500\,K) and interstellar medium absorption lines expected to be present in NUV stellar spectra, the simulator requests the user to also provide the stellar $\ensuremath{\log R^{\prime}_{\rm HK}}$ activity parameter and the Mg{\sc i}, Mg{\sc ii}, and Fe{\sc ii} ISM column densities.\\

\noindent{\bf Transit parameters.} The user can provide a file containing the wavelength-dependent shape of the transit light curve, which is an ascii file where the first column is the time in seconds and the first row is wavelength in Angstrom. The light curve corresponding to each wavelength is then in the corresponding column. If this is not given, the planetary transit shape is set on the basis of the transit impact parameter (in units of stellar radii and ranging between 0 and 1, where 0 corresponds to a central transit), the planetary orbital period (in days), the (wavelength dependent) planetary radius (in stellar radii; i.e., the square root of the transit depth), the orbital inclination axis (in degrees), the orbital semi-major axis (in stellar radii or AU), and the mid-transit time (in Julian Date; JD). If the orbital semi-major axis is larger than one, then the simulator assumes that it is in units of stellar radii (i.e., the actual quantity derived from the analysis of transit light curves), while if it is equal or smaller than 1, then the simulator assumes the semi-major axis to be in AU. If the mid-transit time is not set by the user, the simulator employs the system time in JD. Limb darkening is a further important ingredient setting the transit shape. The simulator employs a quadratic limb darkening law where the limb darkening coefficients in the CUTE band are computed on the basis of {\it PHOENIX}\footnote{{\tt http://phoenix.astro.physik.uni-goettingen.de/}} stellar atmosphere models\cite{husser2013}. The planetary radius can be given either as a single value (i.e., not dependent on wavelength) or as a function of wavelength.\\

\noindent{\bf Target and exposure parameters.} The observing parameters input by the user are the target coordinates RA and DEC (in degrees), the exposure time of a single observation (in seconds), and the time corresponding to the beginning of the first observation (in JD). If this is set to zero and the mid-transit time is not provided, then the simulator takes the time of the first observation from the system time and computes the time of each subsequent observation considering the exposure time and the readout time (see below). If the mid-transit time is set, then the time of each observation is computed on the basis of this value. The total number of consecutive exposures to be computed is automatically set to three times the transit duration.\\

\noindent{\bf Instrument parameters.} These comprise of the position of the star with respect to the slit center (in arcminutes), the spectrograph's spectral resolution at the center of the slit (in \AA), the effective area (in cm$^{2}$) as a function of wavelength (in \AA\ and as an ascii file), the shape of the spectral footprint in the cross-dispersion direction as a function of wavelength (a two dimensional ascii file where the first column corresponds to pixels from the centroid in the cross dispersion direction and the subsequent columns, one for each wavelength bin, give the fraction of the total flux), the slope of the spectrum with respect to the CCD coordinates (i.e., orientation), and the height and width of the slit (in arcseconds).\\

\noindent{\bf Background parameters.} If requested by the user, the simulator can account for both diffuse background and background stars. The parameters (i.e., position, $V$-band magnitude, and $B$-band magnitude) necessary to account for background stars are either taken from the guide star catalog\cite{lasker} or given by the user in an input file. To infer the necessary parameters for the computation of the fluxes of background stars, the simulator employs a look-up table containing spectral type, stellar temperature, $B-V$ color, and stellar radius for main sequence stars taken from Ref.~\citenum{gray2008}. The diffuse background is computed either from the HST STIS high background level\footnote{{\tt http://www.stsci.edu/hst/stis/documents/handbooks/currentIHB/c06\_exptime7.html}} or from zodiacal emission maps\cite{leinert}. Cosmic rays are part of this set of input parameters and the user has to specify the number of cosmic ray hits for any given simulated image. If requested by the user, the simulator further accounts for scattered light, which is controlled by the photon flux in photon\,cm$^{-2}$\,s$^{-1}$\,Sr$^{-1}$, the instrument field of view, and the scattering suppression factor due to multiple reflections, that is the ratio between the number of scattered photons entering the telescope and those reaching the detector.\\

\noindent{\bf CCD parameters.} These comprise the size of the CCD (in pixels), the readout noise (in electrons\,pixel$^{-1}$), the dark level (electrons\,pixel$^{-1}$\,s$^{-1}$), the average bias level (in electrons\,pixel$^{-1}$), the readout time (in seconds), the CCD gain (in photoelectrons\,ADU$^{-1}$), and the instrument's plate scale (in arcseconds\,pixel$^{-1}$). Users can also provide the linearity function of the detector as a two-column ascii file. \\

\noindent{\bf Spacecraft parameters.} These parameters define spacecraft jitter, orientation, observational gaps, and instrument systematics. Spacecraft jitter can be included or excluded and the value set in the simulator corresponds to the RMS jitter in arcseconds\,seconds$^{-1}$ along the x and y directions of the CCD and rotation. The orientation corresponds to the position angle (rotation angle with respect to the RA axis) of the slit, in degrees. Gaps during observations of a given target are due to Earth occultations and the satellite passing through the South Atlantic Anomaly (SAA). To account for gaps, the user has to specify the satellite orbital period and the duration of a typical gap, both in minutes. If requested by the user, the simulator accounts for instrument systematics due defocus caused by temperature variations along a spacecraft orbit, assumed to be of 90\,minutes. The simulator describes this using a third order polynomial, chosen on the basis of what typically observed on HST data (the so-called breathing effect; see for example Ref.~\citenum{sing}, and references therein), for which the user has to provide the coefficients, describing the time-dependent defocus variation.\\

Finally, the user has to set the directories for the location of the input (e.g., stellar models, effective area, etc.) and output files. %{\bf Appendix~\ref{app:} shows an example input parameter file.}
\subsection{Simulator modules and algorithms}\label{sec:algorithm}
\begin{figure}[h!]
\begin{center}
\includegraphics[height=8.1in]{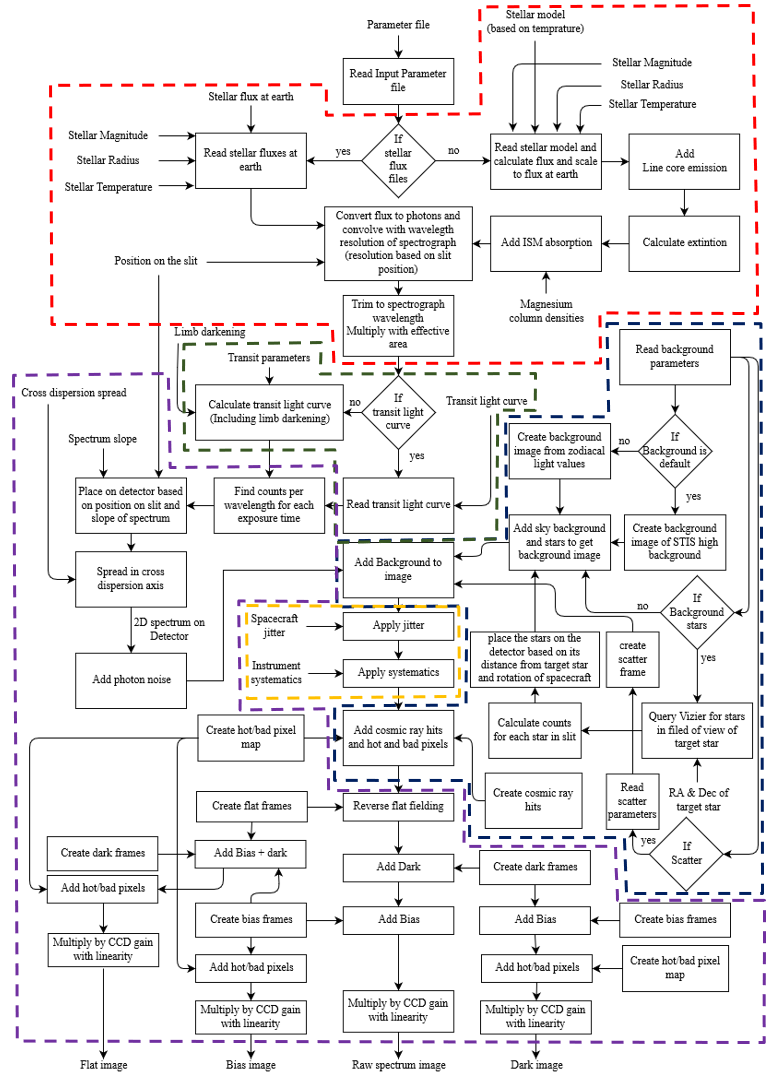}
\caption{Simulator flowchart. The stellar module is contoured in red, the background module in blue, the transit module in green, the spacecraft module in yellow, and the detector module in purple.}
\label{fig:fig1}
\end{center}
\end{figure}

The simulator is structured in separated modules that we thoroughly describe here. The processing steps and logical flow of the simulator are schematically shown as a flowchart in Fig.~\ref{fig:fig1}. The simulator starts by reading the input parameter file and choosing which modules to call on the basis of the input parameters set by the user. 
%-------------------------------------
%-------------------------------------

\noindent{\bf Stellar module.} This module creates the spectrum of the target star and of the background stars starting by either the input file given by the user (only for the target star) or the library of synthetic photospheric stellar fluxes. For the background stars, the temperature is inferred from the $B-V$ magnitude on the basis of the data given in the look-up table. The CUTE band covers also the Mg{\sc ii}\,h\&k resonance lines with chromospheric emission cores in late-type stars with a strength proportional to the activity of the star. To account for this for both target and background stars cooler than 6500\,K, we implemented the emission following Ref.~\citenum{fossati2017}, assuming that the behaviour of the Mg{\sc ii}\,h\&k emission with stellar temperature is the same as that of the Ca{\sc ii}\,H\&K emission, and considering the relation of Ref.~\citenum{linsky} to convert the strength of the disk-integrated Ca{\sc ii}\,H\&K line emission at a distance of 1\,AU ($E$) into Mg{\sc ii}\,h\&k line emission. In this scheme, $E$ depends on the stellar $\ensuremath{\log R^{\prime}_{\rm HK}}$ activity parameter, $B-V$ color, and radius as follows
%-------------------------------------
\begin{equation}\label{eq:E}
E=\frac{(S_{\rm MW}\,10^{8.25-1.67\,BV} - 10^{7.49-2.06\,BV})\,R_{\rm star}^2}{AU^2}\,,
\end{equation}
%-------------------------------------
where $S_{\rm MW}$ is the $S$-index activity indicator in the Mount-Wilson system\cite{noyes1984,mittag2013}, $BV$ is the $B-V$ stellar color, $R_{\rm star}$ is the stellar radius in cm, and $AU$ is one astronomical unit in cm. The exponents in Eq.~\ref{eq:E} are based on calibrations from Ref.~\citenum{mittag2013}, while $S_{\rm MW}$ is defined as
%-------------------------------------
\begin{equation}
S_{\rm MW} = \frac{ 10^{\log R^{\prime}_{\rm HK}} - \frac{10^{7.49-2.06\,BV}}{\sigma T_{\rm eff}^4} }{1.34 \times 10^{-4}\,CF}\,,
\end{equation}
%-------------------------------------
where $T_{\rm eff}$ is the stellar effective temperature and $CF$ is\cite{rutten1984}
%-------------------------------------
\begin{equation}
\log CF = 0.25\,BV^3 - 1.33\,BV^2 + 0.43\,BV + 0.24\,.
\end{equation}
%-------------------------------------
For the target star, the $\ensuremath{\log R^{\prime}_{\rm HK}}$ parameter and radius are given by the user, while the $B-V$ color is inferred from the stellar effective temperature using the look-up table. For the background stars, the stellar radius and temperature are inferred from the $B-V$ color extracted from the guide star catalog and the information given in the look-up table, while the $\ensuremath{\log R^{\prime}_{\rm HK}}$ parameter is set to $-$4.9, which is roughly the average activity level of inactive late-type stars across a wide range of temperatures\cite{fossati2013,staab2017}.

If not provided by the user in the form of a parallax in milliarcseconds, the simulator computes the distance to the star ($d_{*}$) from\cite{fossati2015b}
%-------------------------------------
\begin{equation}\label{eq:distance}
m_{*} = m_{\odot}-2.5\,\log \bigg[\frac{L_{*}}{L_{\odot}}\bigg(\frac{d_{\odot}}{d_{*}}\bigg)^{2}\bigg]
\end{equation}
%-------------------------------------
where $m_{*}$ is the apparent Johnson $V$-band magnitude, $m_{\odot}$ is the apparent Johnson $V$-band magnitude of the Sun (set equal to $-$26.73), $L_{*}$ is the stellar luminosity computed from the input $T_{\rm eff}$ and $R_{\rm star}$ values, $L_{\odot}$ is the solar luminosity, and $d_{\odot}$ is the Sun-Earth distance (i.e., 1\,AU). The stellar flux at Earth is then derived by scaling for the distance. This same operation is carried out for both target and background stars.

The simulator accounts also for insterstellar medium (ISM) extinction and absorption at the wavelength of specific lines covered by the CUTE band. We implemented the extinction for target and background stars using the parametrization by Ref.~\citenum{fitz}, which is valid from the FUV to the infrared (1000\,\AA\ to 35000\,\AA), on the basis of the $E(B-V)$ value obtained from the ISM extinction maps of Ref.~\citenum{amores}, which require a stellar distance (in kpc) and position in the sky (in galactic coordinates), and considering a total-to-selective extinction ratio of 3.05. Guided by STIS UV spectra of DA white dwarfs, we further implemented in the simulator Mg and Fe ISM absorption at the position of the Mg{\sc i} (2852.127\,\AA), Mg{\sc ii} (2795.528\,\AA\ and 2802.705\,\AA), and Fe{\sc ii} (2599.395\,\AA) resonance lines. Also for this we followed the procedure of Ref.~\citenum{fossati2017}. The simulator models each ISM absorption feature as a single Voigt profile with a default value for the $b$-parameter (i.e., broadening) of 3\,km\,s$^{-1}$\cite{malamut} and no radial velocity shift. The Mg{\sc i}, Mg{\sc ii}, and Fe{\sc ii} ISM column densities ($\log N_{\rm ion}$), which set the strength of the ISM absorption features, are either taken from the input parameter file or computed from the ISM abundance and ionisation fraction given by Ref.~\citenum{frisch} (their Table~5) as
%-------------------------------------
\begin{equation}
\log N_{\rm ion} = N_{\rm H} \times {\rm ionization~fraction} \times 10^{\rm abundance}\,,
\end{equation}
%-------------------------------------
where\cite{savage}
%-------------------------------------
\begin{equation}
N_{\rm H} = 5.8 \times 10^{21}\,E(B-V)
\end{equation}
%-------------------------------------
is the total hydrogen column density (i.e., H{\sc i}$+$H$_2$ atoms\,cm$^{-2}$). We took the atomic line parameters for each considered transition (i.e., oscillator strength $\log gf$ and damping constants) from the VALD database\cite{piskunov,kupka,ryab}. This approach of implementing ISM absorption features can be easily extended to any other wavelength with small modifications of the stellar module.

The simulator then converts the stellar flux at Earth, computed as described above, from erg\,cm$^{-2}$\,s$^{-1}$\,\AA$^{-1}$ to photons\,cm$^{-2}$\,s$^{-1}$\,\AA$^{-1}$. In case the user provides the observed stellar spectrum at Earth, the simulator ignores all the steps described above and performs only this conversion. The stellar spectra are then trimmed to the relevant wavelength range (2500--3300\.\AA\ in the case of CUTE) and convolved to the spectrograph's resolution, which depends on the position of the target on the slit in the cross-dispersion direction according to what is given by the ZEMAX ray tracing analysis\cite{fleming}. The ZEMAX analysis indicates that between 2 and 8 arcminutes away from the slit center the resolution degrades to $\sim$66\% of that at the slit center, while the degradation increases to $\sim$50\% at distances from the slit center larger then 8 arcminutes. The simulator then interpolates the trimmed and convolved spectrum to the wavelength scale of the spectrograph and multiplies the result by the effective area of the instrument, that is
%-------------------------------------
\begin{equation}\label{eq:Aeff}
A_{\rm eff}(\lambda) = A_{\rm tot} \times R(\lambda)^{n} \times G(\lambda) \times QE(\lambda)
\end{equation}
%-------------------------------------
to obtain the number of counts\,\AA$^{-1}$\,s$^{-1}$. In Eq.~\ref{eq:Aeff}, $A_{\rm tot}$ is the total collecting area (e.g., in the case of CUTE, this is 122.71\,cm$^2$), $R(\lambda)$ is the wavelength-dependent reflectivity of the optical surfaces, $n$ is the number of reflecting surfaces, $G(\lambda)$ is the grating groove efficiency, and $QE(\lambda)$ is the detector quantum efficiency. For the current version of the simulator, the CUTE effective area is calculated based on the theoretical values provided by the vendors and it will be updated once the instrument is being built and measurements are taken. The simulator then converts the counts\,\AA$^{-1}$\,s$^{-1}$ to counts\,pixel$^{-1}$\,s$^{-1}$ by taking into account the spectral resolution and the number of pixels per resolution element, which in the case of CUTE are 0.8\,\AA\ and 2, respectively.\\

\noindent{\bf Background module.} This module builds the background of the image on which the target spectrum is placed by computing the sky background, scattered light, and the spectra of the background stars in the field of observation. As mentioned above, the user can chose between taking the diffuse background from the high background values of the STIS spectrograph on-board HST or from a zodiacal emission map (see Sect.~\ref{sec:architecture}). Zodiacal light is produced by scattering of solar photons off the interplanetary dust, thus it follows the same spectral shape of the solar spectral energy distribution. The simulator employs the zodiacal emission maps of Ref.~\citenum{leinert}, who tabulated zodiacal light observations as a function of helioecliptic coordinates. Since zodiacal light depends on the position of the Sun and time of the year, the simulator computes the strength of zodiacal emission on the basis of the target position and time of observation.
%-------------------------------------
\begin{figure}
\begin{center}
\includegraphics[scale=0.75]{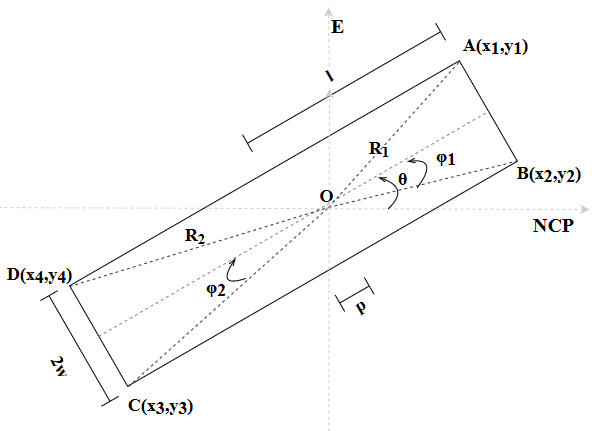}
\caption{Geometry of the slit in the plane of the sky. O corresponds to the position of the target, assumed to lie at the center of the slit in the dispersion direction. $A$, $B$, $C$, and $D$ are the four corners of the slit. The x-axis gives the direction towards the North celestial pole and the y-axis is oriented towards East. $\theta$ is the position angle, $R_{1}$ and $R_{2}$ are the distances from the target star to the corners of the slit, $w$ and $l$ are the half width and half length of the slit in degrees, respectively, and $p$ is the displacement in degrees of the target star from the slit center. For the consideration in this paper we have assumed a slit width of 40" and a length of 20', but the slit size of CUTE is yet to be set.}
\label{fig:fig2}
\end{center}
\end{figure}
%-------------------------------------

The inclusion or not of background stars is chosen by the user through the input parameter file. If the user decides to include background stars, then the simulator queries the guide star catalog\cite{lasker} through ViZieR extracting position (RA and DEC) and $B$- and $V$-band magnitudes for all stars within 12 arcminutes of the target. Once the data from ViZieR have been retrieved, the simulator computes the position of the slit in the sky on the basis of {\it i}) the position of the target star in the sky and in the slit and {\it ii}) the position angle of the slit (i.e., chosen by rotating the spacecraft). Figure~\ref{fig:fig2} shows the geometry of the slit rotated at a position angle $\theta$ in the plane of the sky, where the four edges of the slit, called $A$, $B$, $C$, and $D$ lie at positions ($x_1$,$y_1$), ($x_2$,$y_2$), ($x_3$,$y_3$), and ($x_4$,$y_4$), respectively (see Fig.~\ref{fig:fig2}). From trigonometric rules, it follows that the edges of the slit with respect to the position of the target star, which is at coordinates RA and DEC, are
%-------------------------------------
\begin{equation}
x_{1}={\rm DEC}+R_{1}\,\sin(\theta+\varphi_{1})\,,
\end{equation}
\begin{equation}
x_{2}={\rm DEC}+R_{1}\,\sin(\theta-\varphi_{1})\,,
\end{equation}
\begin{equation}
x_{3}={\rm DEC}+R_{2}\,\sin(180+\theta+\varphi_{2})\,,
\end{equation}
\begin{equation}
x_{4}={\rm DEC}+R_{2}\,\sin(180+\theta-\varphi_{2})\,,
\end{equation}
\begin{equation}
y_{1}={\rm RA}+R_{1}\,\cos(\theta+\varphi_{1})\,,
\end{equation}
\begin{equation}
y_{2}={\rm RA}+R_{1}\,\cos(\theta-\varphi_{1})\,,
\end{equation}
\begin{equation}
y_{3}={\rm RA}+R_{2}\,\cos(180+\theta+\varphi_{2})\,,
\end{equation}
and
\begin{equation}
y_{4}={\rm RA}+R_{2}\,\cos(180+\theta-\varphi_{2})\,,
\end{equation} 
%-------------------------------------
where $\theta$ is the position angle and $R_{1}$ and $R_{2}$ are the radial distances from the target star to the corners of the slit, which are given by
%-------------------------------------
\begin{equation}
R_{1}=\sqrt{w^{2}+(l-p)^{2}}
\end{equation}
and
\begin{equation}
R_{2}=\sqrt{w^{2}+(l+p)^{2}}\,.
\end{equation}
%-------------------------------------
Here $w$ and $l$ are the half width and half length of the slit in degrees, respectively, and $p$ is the displacement in degrees of the target star from the slit center (see Fig.~\ref{fig:fig2}). The angles $\varphi_{1}$ and $\varphi_{2}$ are given by
%-------------------------------------
\begin{equation}
\varphi_{1}=\arctan\left(\frac{w}{l-p}\right)
\end{equation}
and
\begin{equation}
\varphi_{2}=\arctan\left(\frac{w}{l+p}\right)\,.
\end{equation}
%-------------------------------------
In the above equations, all angles are given in degrees. Once the position of the slit on the sky is known, the simulator checks which background stars lie inside the slit and for those computes their x and y position on the CCD on the basis of the angular distance between the star and the direction of the star with respect to the target star as 
%-------------------------------------
\begin{equation}
x=d\,\sin{\alpha}
\end{equation}
and
\begin{equation}
y=d\,\cos{\alpha}\,,
\end{equation}
%-------------------------------------
where $d$ and $\alpha$ are
%-------------------------------------
\begin{equation}
d=\sqrt{({\rm RA}_{s}-{\rm RA})^{2}+({\rm DEC}_{s}-{\rm DEC})^{2}}
\end{equation}
and
\begin{equation}
\alpha=\arcsin\left({\frac{{\rm RA}_{s}-{\rm RA}}{d}}\right)-\theta\,,
\end{equation}  
%-------------------------------------
where RA$_{s}$ and DEC$_{s}$ are the coordinates of the particular background star. Any star whose position is within one arcsecond of the target is excluded to avoid the spectrum of the target star being reproduced twice, thus implicitly assuming that there are no other stars within one arcsecond of the target. We remind the reader that CUTE has a plate scale of 2.5 arcseconds, thus making any target within this radius indistinguishable. The stellar module then generates a spectrum for each relevant background star and shifts it relative to its position in the sky. We remark that the simulator carries out the shift along the x-axis both in physical pixels and wavelength.

The background module also produces an image of the field of observation (set as 12 arcminutes radius from the target star) with the slit superimposed, taking the position angle into account (Fig.~\ref{fig:fig9}). Each circle in Fig.~\ref{fig:fig9} represents a star with the circle's size being proportional to the stellar $V$-band magnitude, as indicated by the legend, while an `x' indicates the position of a star with an unknown $V$-band magnitude. These maps serve as finding charts that can be used to set up the orientation of the spacecraft (i.e., of the slit) in the sky to minimize contamination from nearby stars. 

We model scattered light as a constant source of light distributed uniformly across the detector. The simulator takes the total amount of photons incident on the telescope and coming from directions other than the direction of pointing and scales it for the slit field of view and the depletion factor, which are all provided by the user. The scattered photons reaching the detector are then uniformly spread across it. This scattered light model assuming a constant distribution will be replaced by the actual scatter model for the instrument; this analysis is still ongoing.

Airglow emission from the Earths atmosphere is a relevant source of contamination for ultraviolet space missions in low earth orbit, particularly at wavelengths shorter than those probed by CUTE. A comprehensive review for airglow is given by Ref. \citenum{meier}. Figure 10 of reference \citenum{meier} is an absolute worst case representation of CUTE observations (satellite orbiting only at an altitude of 184 km and not pointing away from the Earth) and even in that case airglow only accounts to at worst one count per 300 second exposure per Angstrom. It will be much less for a typical CUTE observation due to the fact that we will be observing from a much higher altitude and will be observing always away from the Earth. This and the fact that the Earth atmosphere does not produce significant airglow emission at the wavelengths probed by CUTE explains why we did not implement airglow contamination in the simulator.

\noindent{\bf Planetary transit module.} This module implements a time- and wavelength-dependent variation to the flux of the target star according to the input parameters of the planetary transit, namely transit duration and wavelength-dependent shape and depth. The complete simulated time-span corresponds to three times the duration of the planetary transit, where the transit is placed in the middle of the considered simulated time, and with a time resolution of 0.005\% of the transit duration. The simulator computes the shape of the wavelength-dependent transit light curve on the basis of the user input and the transit shape computed with EXOFAST\cite{eastman}, employing the planetary system parameters given by the user and the wavelength-dependent quadratic limb-darkening coefficients computed from PHOENIX synthetic spectra\cite{husser2013} in 10\,\AA-wide wavelength bands. We have only considered limb darkening and not limb brightening because limb brightening would be relevant just for the emission in the line core of  the MgII h\&k lines, which CUTE is unlikely to resolve. Because of the low flux at the core of the MgII h\&k lines, the light curves covering MgII will always include a significant portion of the line wings. Furthermore, studies of limb brightening for the MgII lines as a function of stellar mass, temperature, and activity are not available. The parameters that need to be set in the input parameter file for the computation of the transit are the location of the library of PHOENIX models, the orbital period ($P_{\rm orb}$), semi-major axis ($a$), and inclination angle ($i$), the impact parameter ($b$), and the stellar and planetary radii ($R_{\rm star}$ and $R_{\rm pl}$). From these value, the simulator computes the transit duration ($T_{\rm dur}$) as
%-------------------------------------
\begin{equation}
T_{\rm dur}=\frac{P_{\rm orb}}{\pi}\,\arcsin\left(\frac{\sqrt{(R_{\rm star}+R_{\rm pl})^2-b^2}}{a\,\sin i}\right)\,.
\end{equation}
%-------------------------------------
Once the shape of the transit light curve has been computed, the simulator locates the time of each exposure, taking into account the readout time, and computes the average transit depth at each given exposure, which is then used to modulate the flux of the target star as a function of time.\\

\begin{figure}
\begin{center}
\includegraphics[width=\textwidth]{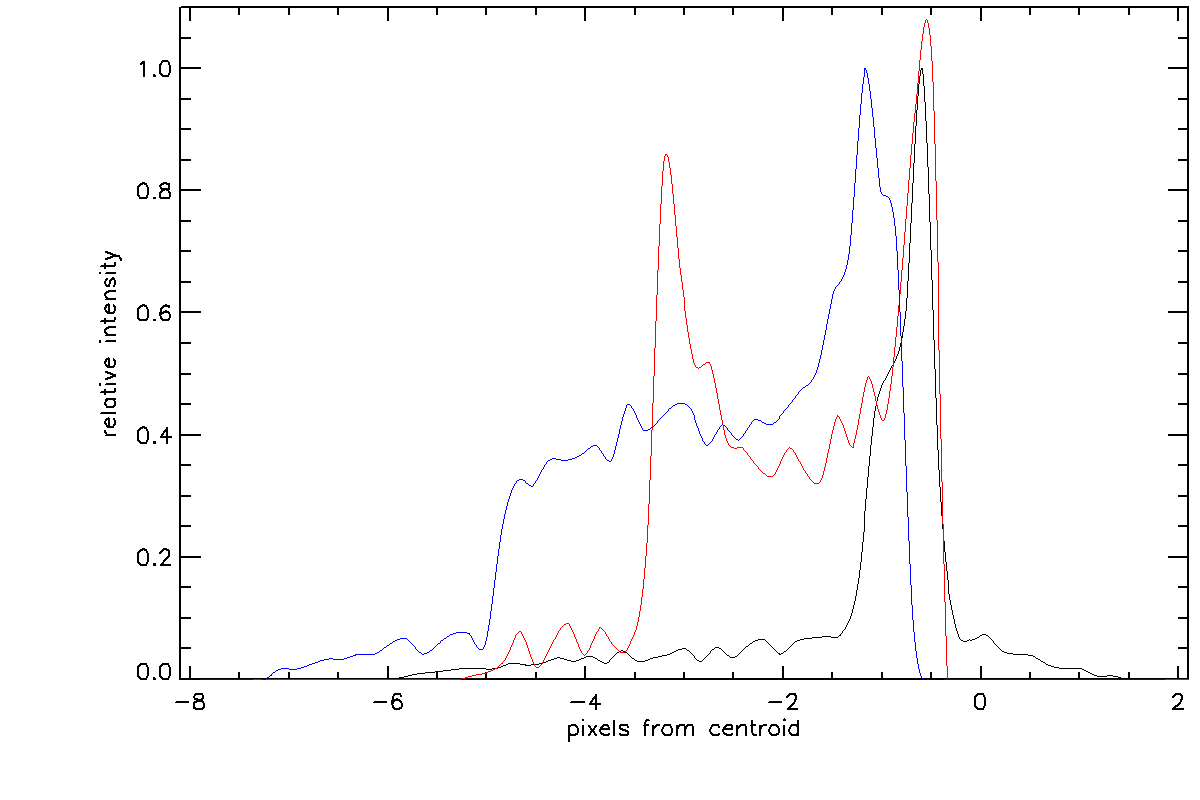}
\caption{Shape of spectrum along the cross-dispersion direction derived from ZEMAX at the center of the detector (i.e., 2900\,\AA; red) and at the edges (2550\,\AA; black - 3300\,\AA; blue).}
\label{fig:fig3}
\end{center}
\end{figure}

\noindent{\bf Detector module.} This module creates the two-dimensional spectrum of the target and background stars on the CCD, accounting for all detector noise sources (e.g., bias, flat) and cosmetics (e.g., cosmic rays). CUTE uses an Teledyne e2V CCD42-10 rectangular 2048$\times$515 pixels CCD detector, with a pixel size of 13.5\,$\mu$m. This is an UV-enhanced back-illuminated detector with an active area of 27.6$\times$6.9\,mm. This was the detector of choice on the Mars Science Laboratory ChemCham LIBS spectrometer and hence has proven flight heritage. In order to keep the dark noise low the CCD will be cooled to Peltier temperatures around -$50^\circ$\,C. The resolution element of the CUTE spectrograph is two pixels. This module spreads the spectrum along the cross-dispersion axis. We obtained the shape of the spectrum in the cross-dispersion axis from a ray-tracing analysis conducted with ZEMAX. We find that, because of optical aberrations, the shape of the spectrum along the cross-dispersion axis is non-Gaussian and varies across the detector. To account for this variation, though maintaining the computation relatively simple, we divided the whole spectral coverage of about 800\,\AA\ into 10 segments, assigning a different shape to the spectrum in the cross-dispersion direction to each of these segments. Figure~\ref{fig:fig3} shows the shape of the spectrum in the cross-dispersion direction at the center and edges of the CCD. Even though the current system is tailored to CUTE, this module enables the user to change the input file describing the shape of the spectrum in the cross-dispersion direction, thus adapting it to other instruments, though always keeping the assumption of 10 segments across the whole spectral coverage. For general purpose, we have also build in the possibility to use a Gaussian shaped spectrum in the cross-dispersion direction. This module also applies the spectral slope along the x-axis of the detector provided by the user. The detector module creates the image frame by defining an array of the size of the detector and placing the spectra of the target and background stars on the CCD according to their position on the sky and relative to the slit, considering a sub-exposure time of 1\,second to further account for spacecraft jitter (see spacecraft module).
%-------------------------------------
%-------------------------------------

\begin{figure*}[ht!]
\centering
\begin{center}
\includegraphics[width=\textwidth]{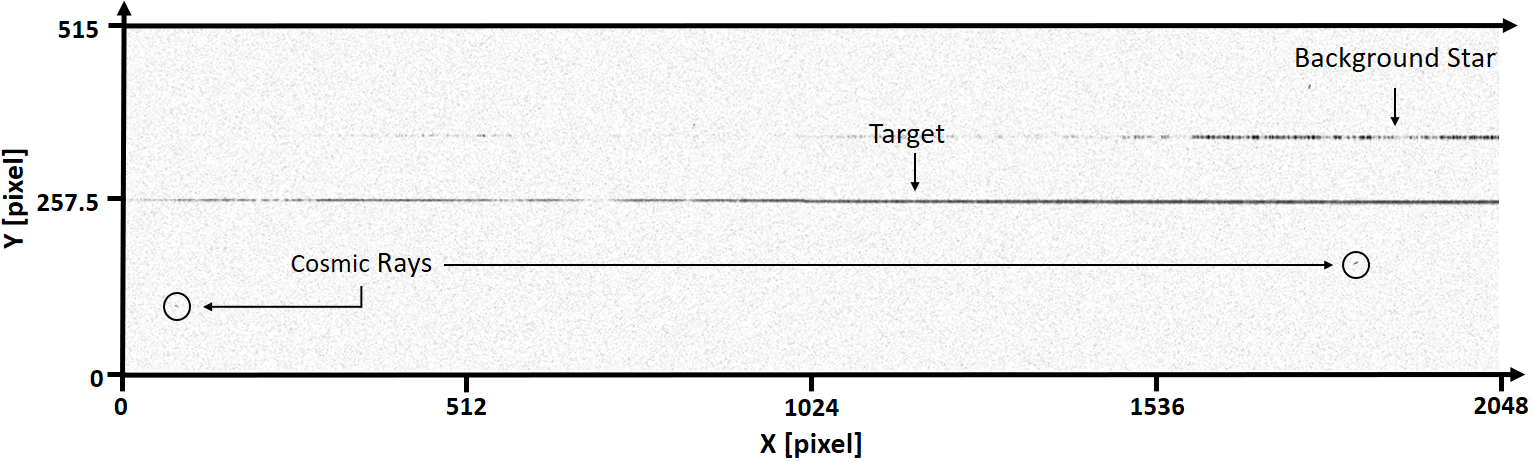}
\caption{Example raw image of the planet-hosting star KELT-7 produced by the simulator. The figure highlights cosmic ray hits and the only clearly visible background star.}
\label{fig:fig4}
\end{center}
\end{figure*}

Random cosmic ray hits are a source of noise in any astronomical observation. In low-Earth orbit, the typical number of cosmic rays is\footnote{\tt www.stsci.edu/instruments/wfpc2/Wfpc2\_hand/wfpc2\_handbook.html} 1.2 events\,s$^{-1}$\,cm$^{-2}$, which is the value we set as default, but it can be changed by the user. Cosmic ray hits on the CCD can generate a sequence of consecutive pixels to saturate. To simulate this effect, this module sets the number of counts equal to the saturation level for a sequence of five consecutive pixels randomly placed and oriented across the CCD. Figure~\ref{fig:fig4} shows a CUTE CCD image generated with the simulator, highlighting two cosmic rays.

This module also adds photon noise to the stellar spectra where the magnitude of the noise is the square root of the number of photons per pixel and applies a non-linearity correction to each pixel based on an input file. If an input file describing the linearity of the CCD is not specified, the gain is set to remain constant (i.e., linearity regime) up to 60\,000\,ADU, above which the gain follows a third order polynomial until the saturation level of 72\,000\,ADU. As part of this module, the simulator creates five bias, dark, and flat-field frames. The user gives the average bias level in the input parameter file and the simulator creates five different bias frames with pixel-to-pixel variations taken from a random generator considering a normal distribution with a sigma equal to the readout noise, which is also given as input by the user. The simulator employs the average dark level in counts\,pixel$^{-1}$\,s$^{-1}$ given by the user to generate the average dark frame given a certain exposure time and generates five different dark frames in the same way as for the bias frames, but with a noise level equal to 0.1\% of the average dark value. The simulator creates an ideal flat field, which is constant across the detector and set to 50\,000\,ADU. The simulator creates then five flat field images following the same procedure as for the bias and dark images, but considering a noise level of 5\,ADU, that is assuming a signal-to-noise of the flat field of 10\,000. We will substitute the bias, dark, and flat field frames generated by the simulator with the actual frames as soon as they become available as part of the characterisation campaign of the CUTE detector\cite{egan}. This module gives the user the additional possibility to add a hot/bad-pixel mask to account for cosmetic imperfections of the CCD.\\

\noindent{\bf Spacecraft module.} Depending on the quality of the attitude control system, the pointing of each spacecraft is affected by jitter, which for CUTE is expected to be smaller than 7.2\,arcseconds RMS per second. These small high-frequency pointing variations cause the stars to move randomly within the slit, thus slightly reducing the spectral resolution. To model the effect of jitter on the output spectra, the user has to provide in the input parameter file the RMS value of jitter in arcseconds\,s$^{-1}$ in the three perpendicular directions. Let's now consider the x and y axes to lie respectively along the dispersion and cross-dispersion axes of the CCD and the z axis to lie in the direction perpendicular with respect of the plane of the CCD. The simulator then assumes the focal plane of the instrument to lie in the x-y plane, thus rotations in yaw (around the x-axis) and pitch (around the y-axis) angles translate into linear displacements in pixels in x and y on the CCD. Jitter in the roll angle (around the z-axis), which is perpendicular to the focal plane, rotates instead the image on the CCD. To mimic jitter, the simulator splits a single exposure into several one-second sub-exposures and applies to each of them a random translation and rotation having an RMS amplitude corresponding to that of the jitter in the three axes. The sub-exposures are then co-added to generate an image of the given exposure time.

Once the effect of spacecraft jitter has been added to the spectra of the target and background stars, the simulator adds to the image the effects of bias, dark, and flat field and the resulting image is finally multiplied by the CCD gain to get the final raw frame. Together with the raw frames, the simulator produces the calibration images (bias, dark, and flat field), and images containing the spectrum of the target star without noise and spacecraft jitter. Figure~\ref{fig:fig4} shows an example raw image produced by the simulator.
%-------------------------------------
%-------------------------------------

Targets are usually distributed uniformly across the sky meaning that in many cases the Earth will occult a targets for a certain period of time along each spacecraft orbit. To account for this, the simulator randomly places a gap in observation of the duration specified by the user and starting within a time corresponding to the duration of the spacecraft orbit following the first observation. The gap then repeats itself with a periodicity equal to the duration of the spacecraft orbit.

This module applies also the effect of instrumental systematics to the sequence of spectra taken across a spacecraft orbit. We consider the instrument systematics to be tied to focus variations caused by temperature fluctuations along an orbit, with the variations reproducing themselves at each orbit. The simulator models systematics as a time-dependent third order polynomial reproducing the focus fluctuations as a function of spacecraft orbital phase. This is similar to the breathing effect present in HST data. Because of the anticipated Sun-synchronous, dawn-dusk orbit of CUTE, we expect no significant temperature fluctuations, thus negligible systematics. However, this feature of the simulator can become useful in case CUTE will end up on a different orbit and for other missions with different orbits.\\

\noindent{\bf Quick data reduction.} To enable a quick analysis of the raw images, we added to the simulator a quick data reduction pipeline (QDP), which can be executed separately from the simulator. Following bias and dark subtraction and flat fielding removal, the QDP extracts the spectrum and the background within areas chosen by the user. Once the spectrum of the target star has been extracted, the QDP converts it to photons, by dividing by the effective area, and then, optionally, flux calibrated, if the required calibration information is given by the user. Finally, the QDP converts the pixels to wavelength using an ascii  file given by the user. The QDP does not correct for cosmic rays and the extraction does not account for background stars, thus the pipeline is designed mostly to provide a quick way to verify the output of the simulator. The QDP is however the baseline on which we will build the CUTE data reduction pipeline, which will be presented in a separate work.

%\section{CUTE simulator software package}
%%-------------------------------------
%The simulator, ACUTEDIRNDL\footnote{Available at: {\tt https://github.com/agsreejith/ACUTEDIRNDL}}, set of IDL routines is governed by an input parameter file called {\it cutedrndl\_parameters.txt} available with the software package. The simulator can be run after editing the parameter file. The outputs will be stored in the folder specified in input parameter file. The number of output files will depend on both transit duration and exposure time per observation. By default the simulator overwrites the files in the output directory. The detailed description of the parameters in the input parameter file is described in section 2.1. 
%
\section{Results}\label{sec:results}
We provide here a few practical examples of application of the simulator to CUTE aiming at a better understanding of the mission's data products and their characteristics. We remark that, being based on realistic, though still simulated, calibration frames, the results shown here have to be taken as an indication of the actual future capabilities of CUTE and are not exact. We will present more up to date results together with the data reduction pipeline.
\subsection{Transit light curve of the typical hot Jupiter HD209458b}
%

%-------------------------------------
\begin{figure}[hb]
\begin{center}
\includegraphics[width=\textwidth]{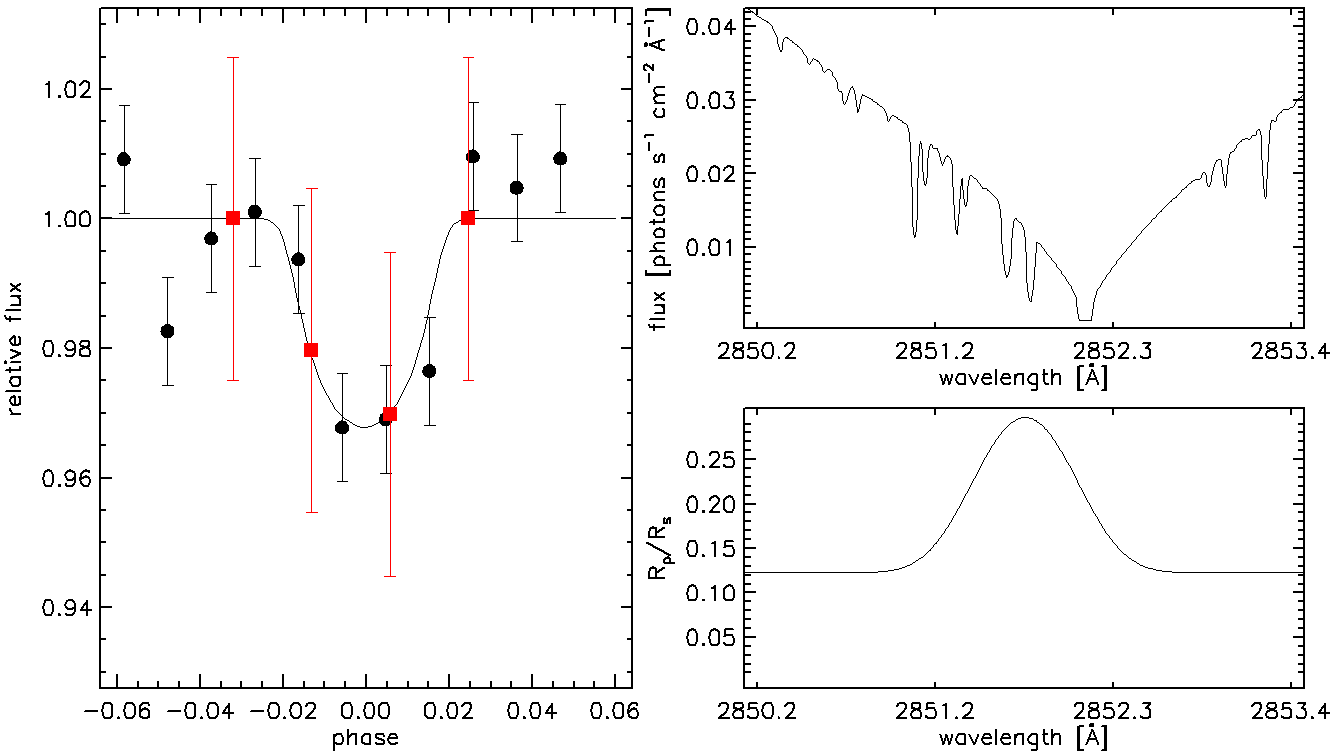}
\caption{Left: HD209458b transit light curve in a four resolution element (3.2\,\AA\ or 337\,km\,s$^{-1}$) bin around the Mg{\sc i} 2852\,\AA\ resonance line. The light curve (black filled circles) was obtained combining ten transits and by rebinning over ten consecutive CUTE frames. The red squares show the data-points and uncertainties presented by Ref.~\citenum{vidal} over a 50\,km\,s$^{-1}$ bin. By accounting for the difference in the considered integration intervals between HST (50\,km\,s$^{-1}$) and CUTE (337\,km\,s$^{-1}$), the uncertainties on each data point obtained from HST and expected for CUTE are about the same. Top-right: stellar spectrum in the spectral region considered for the integration. Bottom-right: input transmission spectrum in the spectral region considered for the integration.}
\label{fig:fig5}
\end{center}
\end{figure}
%-------------------------------------

As an example, we present here the simulated transit light curve of the typical hot Jupiter HD209458b as observed with CUTE. Because of the brightness of the host star, of the rather large transit depth, and of the presence of STIS near-UV transit observations\cite{vidal} to be used as direct comparison, the HD209458 system is a primary CUTE demonstration target. Figure~\ref{fig:fig5} shows the CUTE transit light curve obtained combining together the results of ten transits and integrating over four resolution elements around the core of the Mg{\sc i} 2852\,\AA\ resonance line and assuming no time variability in the intrinsic stellar flux and planetary radius. This is the line for which Ref.~\citenum{vidal} detected a transit depth of about 7\% using HST STIS integrating over about 50\,km\,s$^{-1}$ close to the line core. To set up the simulation, we have taken the transmission spectrum around the Mg{\sc i} 2852\,\AA\ resonance line presented by Ref.~\citenum{vidal} and fit a Gaussian profile to it, which is then used as the input transmission spectrum (see bottom-right panel of Fig.~\ref{fig:fig5}). Because the actual orbit of the satellite is still unknown, we did not include gaps in the simulation.

%-------------------------------------
\begin{figure}[ht]
\begin{center}
\includegraphics[width=\textwidth]{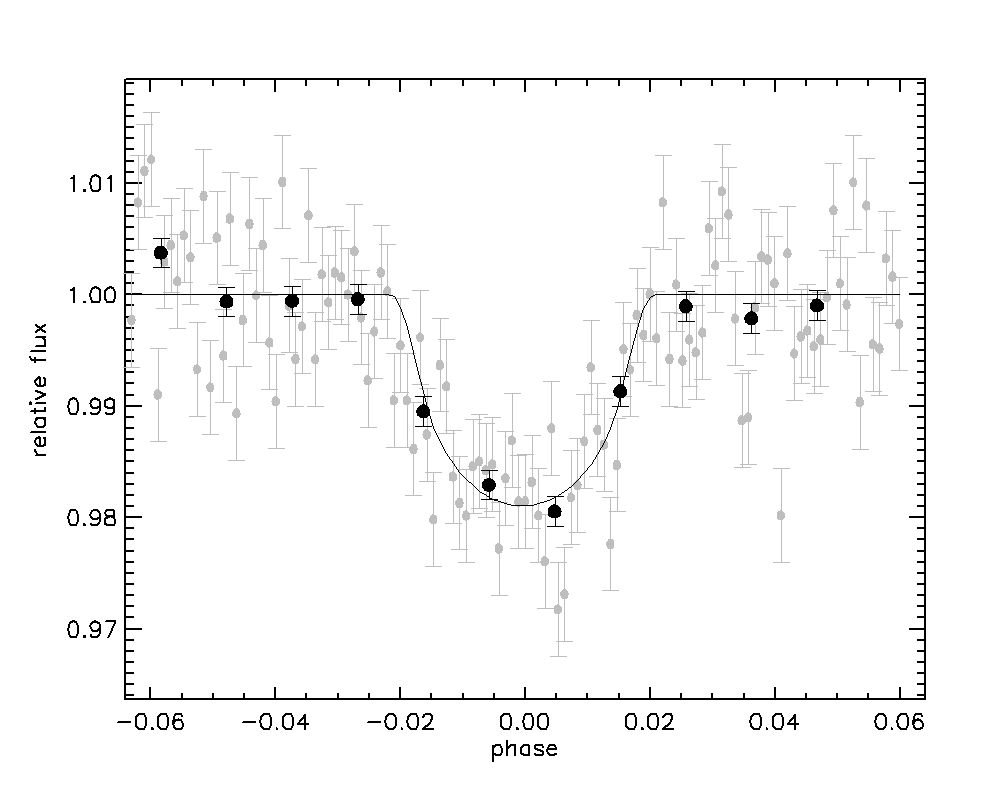}
\caption{HD209458b single transit light curve obtained integrating over a 100\,\AA\ continuum region (2884--2984\,\AA) and re-binning (averaging) ten consecutive frames.}
\label{fig:fig6}
\end{center}
\end{figure}
%-------------------------------------

Figure~\ref{fig:fig5} shows that within ten transits CUTE will be able to clearly detect at high significance the extra-absorption caused by the presence of Mg{\sc i} in the planetary upper atmosphere, with uncertainties comparable to those obtained with HST within a single transit, thus meeting the mission's basic requirements. This plot further shows that CUTE allows one to uniformly sample the whole transit, which is difficult to reach with HST because of the usually limited number of observed transits per system. Figure~\ref{fig:fig6} shows a further simulation where the signal is extracted over $\sim$100\,\AA\ of continuum emission for a single transit. This shows that CUTE will be able to detect typical hot-Jupiter transit signals even within one transit observation.  
\subsection{Precision of CUTE transit observations as a function of temperature and magnitude of the host star}
\begin{figure}[h]
\begin{center}
\includegraphics[width=\textwidth]{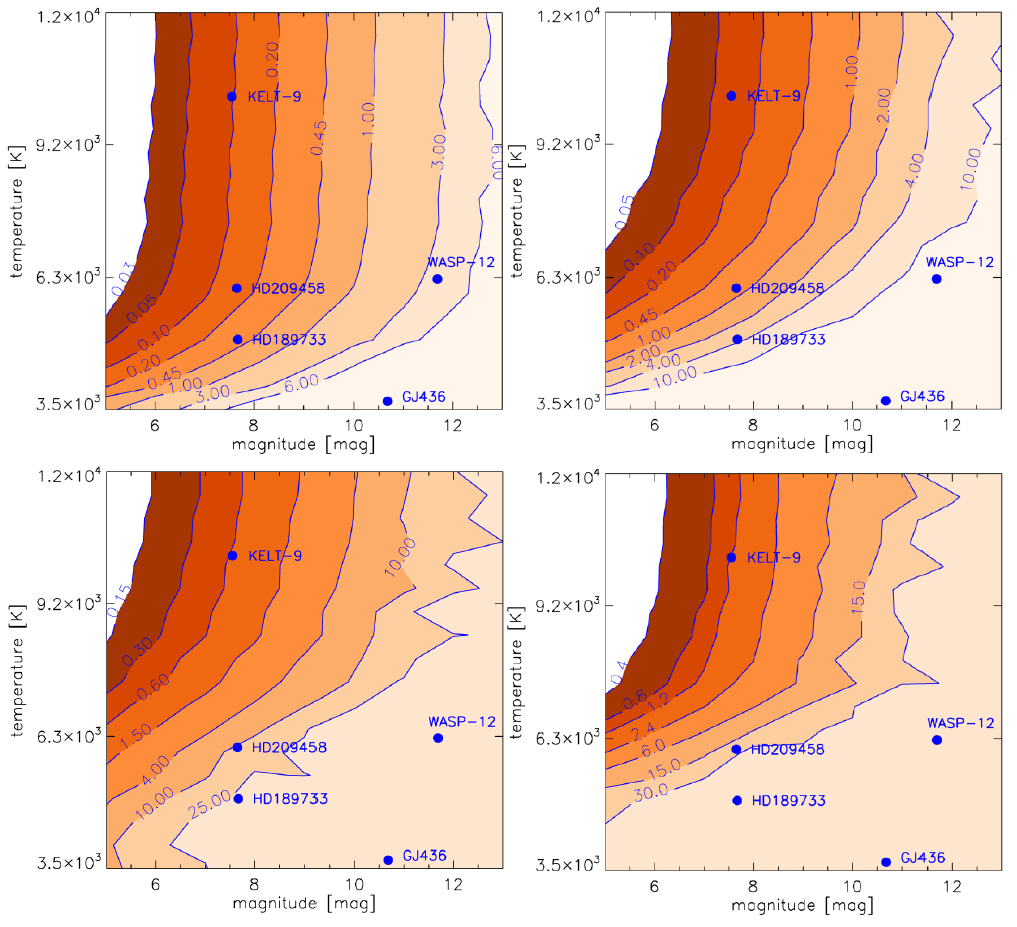}

\caption{Top-left: Uncertainty on the transit depth, in \%, for a 5\,minutes CUTE observation integrating in wavelength above 3000\,\AA. Top-right: as top-left, but integrating in wavelength below 2750\,\AA. Bottom-left: same as top-left, but integrating in wavelength around the Mg{\sc ii}\,h\&k resonance lines (2790--2810\,\AA). Bottom-right: same as top-left, but integrating in wavelength around the Mg{\sc i} resonance line (2850--2854\,\AA). Each panel marks also the position of systems for which signatures of atmospheric escape have been observed in the past\cite{vidal2003,fossati2010,lecavelier2012,ehrenreich2015,yan2018}.These maps have been derived considering the effect of jitter.}
\label{fig:matrix}
\end{center}
\end{figure}

We employed the simulator to estimate the precision on the transit depth, in \%, that will be obtained with CUTE by integrating over four different wavelength regions as a function of magnitude and effective temperature of the host star (see Fig.~\ref{fig:matrix}). Two of the four selected wavelength ranges are broad and cover the region with the highest stellar flux in the CUTE band (i.e., above 3000\,\AA) and the region below 2700\,\AA, which has been shown to be also sensitive to exoplanet atmospheric escape\cite{salz2018}. The other two regions are centered around the Mg{\sc ii}\,h\&k resonance lines and the Mg{\sc i} resonance line at 2852\,\AA. The former band is comparable to one of the NUV wavelength bands covered by COS for the transit observations of WASP-12b\cite{fossati2010,haswell2012}.

We have run the simulator for stellar effective temperatures varying between 3500 and 12000\,K in steps of 500\,K and $V$-band magnitudes varying between 5 and 13\,mag, in steps of 1\,mag. For the computations we have considered a stellar radius varying with temperature following the look-up table, a $\ensuremath{\log R^{\prime}_{\rm HK}}$ parameter for the late-type stars of $-$4.9, an exposure time of 300\,seconds, a read-out time of 60\,seconds, no extinction, and an RMS jitter of 7.2\,arcseconds\,s$^{-1}$, and excluded background stars and cosmic rays. The distance to each star was computed by the simulator from the information given in the input parameter file and using Eq.~\ref{eq:distance}.

%-------------------------------------
%-------------------------------------

As expected, Fig.~\ref{fig:matrix} shows that the uncertainty on the transit depth decreases with increasing stellar temperature and decreasing magnitude. Considering an average transit duration of 2.5\,hours, without gaps over one transit the precision shown in Fig.~\ref{fig:matrix} improves by a factor of about five. At the wavelengths of lines probing atmospheric escape, transit depths are typically larger than 2--3\%, therefore Fig.~\ref{fig:matrix} shows that within a few transits CUTE will be capable of detecting escape for planets orbiting stars brighter than the 13th magnitude and hotter than about 6500\,K. For cooler stars, the detection of atmospheric escape will be limited to the brighter ones, such as HD189733, for which the necessary precision will require the observation of about 10 transits. The case of KELT-9 is particularly remarkable as within a single transit CUTE will be able to reach a precision on the transit depth at the position of the Mg{\sc i} and Mg{\sc ii} lines of about 0.1\%. This translates to a precision on the planetary radius of about 2.2\%, which corresponds to about two pressure scale heights. Scattered light can affect the noise level for faint targets, however the noise budget for faint stars is significantly dominated by readout noise, which is why scattered light does not play a role, even in the worse case scenario of a scattered light comparable to the sky background.

%%
%\subsection{Tests on noise levels}
%%
%Other than photon noise, instrumental noise sources include the noises associated with the detector and readout electronics including readout noise, dark noise and bias values. To better understand the effect of detector noise and the pixel to pixel variation in the transit light curve at various wavelengths, we have carried out the following exercises. Dark noise level and readout noise level where varied independently. All other input parameters were kept constant. A linear variation of readout noise in input creates a similar linear change in noise level on the output. It was also observed that of all the noise sources in the instrument the readout noise dominates and plays the role of limiting the signal to noise ratio of the observations, due to extremely low dark noise of 1.2$\cdot$ 10$^{-2}$ e pixel$^{-1}$ s$^{-1}$ obtained by a cooled detector (CUTE detector is planned to be operated at around -50$^{\circ}$C\cite{fleming}.  
%
\subsection{Effect of spacecraft jitter on the spectral resolution}
\begin{figure}[h]
\begin{center}
\includegraphics[width=\textwidth]{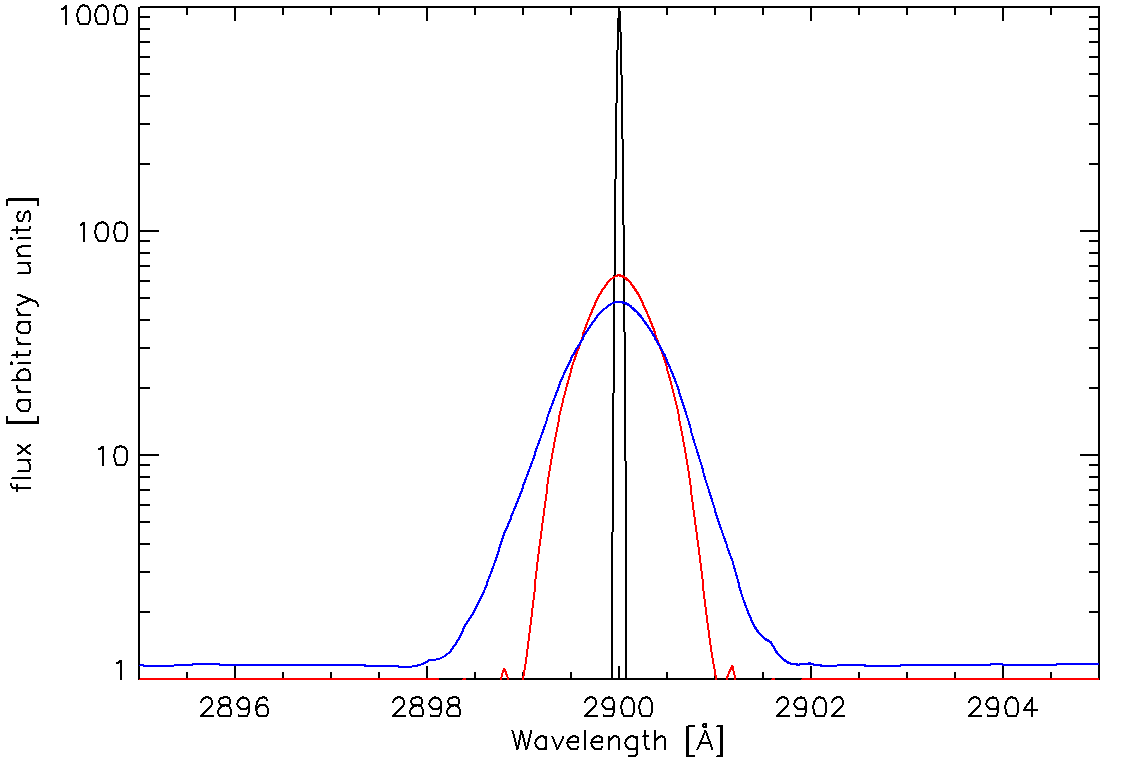}
\caption{Gaussian profiles given as input to the simulator (black) and obtained after a full simulation with (blue) and without (red) spacecraft jitter.} 
\label{fig:fig8}
\end{center}
\end{figure}

Random movements of the star on the slit due to spacecraft jitter degrade the final spectral resolution. The upper limit on the pointing accuracy of the XB1 spacecraft, as listed by the provider (Blue Canyon Technologies), is 7.2\,arcseconds\,s$^{-1}$ RMS \cite{fleming}. To understand the effect of jitter on the final spectral resolution of CUTE data, we generated an emission line spectrum composed of a single Gaussian centered at 2900\,\AA\ and with a FWHM of 0.05\,\AA, which is much smaller than the nominal resolution of the CUTE spectrograph (0.8\,\AA). We have then passed this spectrum to the simulator running it with and without jitter. Without jitter, we obtained a Gaussian line profile with the expected width of 0.8\,\AA. With jitter, instead, the width of the Gaussian line profile increases to 1.2\,\AA, which is still within the requirements\cite{fleming}. Figure~\ref{fig:fig8} shows the three Gaussian profiles. We further employed this experiment to test how well the simulator conserves flux throughout the different steps, obtaining that flux is conserved within 0.003\%, which is what one would expect considering edge trimming during the convolution and machine accuracy.
%-------------------------------------
%-------------------------------------

The CUTE instrument team is looking into options to decrease or control spacecraft jitter, thus improving the final spectral resolution aiming possibly at bringing it below 5\,arcseconds RMS over long timescales. One option is to favour jitter with a periodicity longer than the typical CUTE exposure time of 5\,minutes over short-period jitter. %{ \bf to remove: Another option is to favour jitter along the cross-dispersion direction rather than along the dispersion direction, thus decreasing the detrimental effect of jitter on the final spectral resolution.}
%Our team at the University of Colorado is looking into options for constraining or controlling jitter, so as to favor one form of jitter for eg., slow period jitter with timescales greater than the CUTE exposure time over high-frequency jitter, or the possibility of favoring jitter in one axis over another (spatial instead of spectral, or vice versa), which may get the effective jitter below 5". Apart from spectral broadening the signal to noise ratio is also affected by jitter. This noise will play an important role if the spectral extraction is carried out only over a small wavelength range (for eg. a specific atomic or molecular species).
%
\subsection{Spacecraft orientation}
\begin{figure*}[h]
\begin{center}
\includegraphics[width=\textwidth]{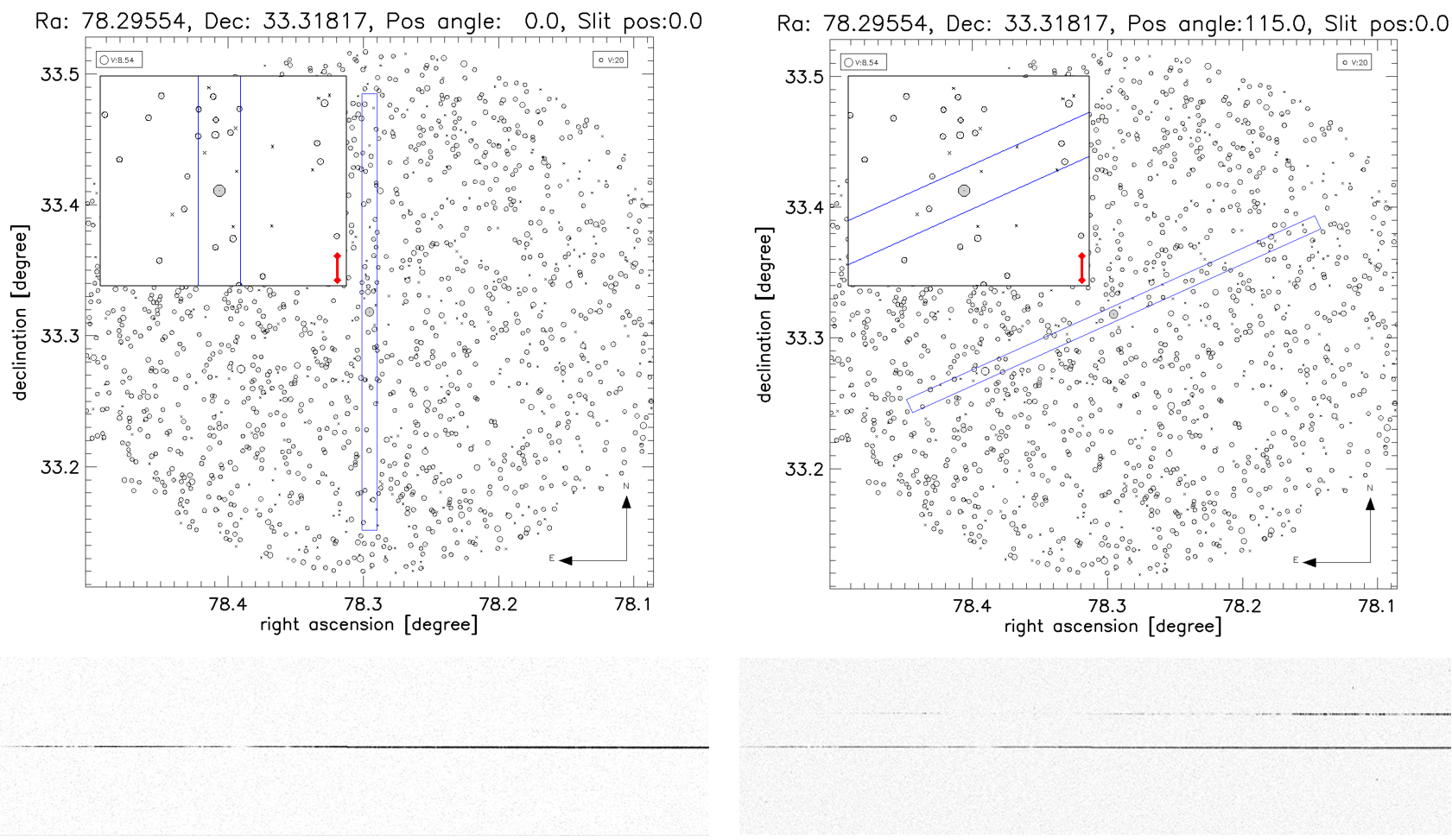}
\caption{Top: maps of the CUTE field of view centered at the position of KELT-7 with two different oriantations of the slit, which is shown by the blue rectangle. The symbol size is proportional to the stellar $V$-band magnitude as given by the guide star catalog. Stars marked by an `x' do not have a $V$-band magnitude listed in the catalog. The slit is 20\,arcminutes long and 40\,arcseconds wide. The central region around the target star is shown in the top-left inset (230"$\times$200") for better clarity. The red line on the bottom-right of the inset shows the extent of 10 pixels. Bottom: raw CUTE images obtained with the simulator considering the two different orientations of the slit.}  
\label{fig:fig9}
\end{center}
\end{figure*}

Choosing the right orientation of the spacecraft can be the key to mitigate the problem of blending from background stars for targets in crowded fields, such as HD189733 and KELT-7. The map of the field of view of the simulated target aims at allowing the user to best select the orientation of the simulator. In addition, these maps and the results of the simulation allow the user to pre-define the extraction regions for the target star and the background to prevent blending. Figure~\ref{fig:fig9} presents the map of the field of view and one raw image for KELT-7, which lies in a particularly crowded field, for two different orientations of the spacecraft.
%-------------------------------------
%-------------------------------------
%
\section{Summary and future work}\label{sec:conclusions}
We presented the structure of the CUTE data simulator and the algorithms implemented in it. Its modular framework enables one to adapt this tool to simulate data produced by other spacecrafts carrying on-board a long-slit spectrograph and a CCD detector. We plan to employ the simulator to generate the data required to set up the CUTE signal-to-noise calculator and to develop the CUTE data reduction pipeline. We will further use the simulator to generate synthetic transit light curves for systems in the CUTE target list to aid the target selection. The simulator has been built in such a way to also enable the user to input any wavelength-dependent transit shape, thus allowing us to produce simulated data in the presence of asymmetric transits\cite{fossati2010}, hence allowing the science team to test the capabilities of CUTE and of the data analysis tools to detect these features that are one of the main mission's science drivers.

By the time of CUTE's launch, we plan to implement a number of additional features in the simulator that will make it more accurate, versatile, and easy to use. 
\begin{itemize}
\item Following the characterization campaign of the CUTE CCD, we will substitute the part of the detector module generating the bias, dark, and flat field frames with the actual images obtained directly from the flight CCD.
\item At present, the spread in the cross-dispersion direction is defined in ten different wavelength bands, but we will significantly improve the mapping of the shape of the spectrum across the whole detector, possibly going to the sub-pixel level to avoid jumps.
\item The sub-exposure for the implementation of jitter is currently set to one second, but we will make it a user input. This will enable the user to trade between the speed of the computation and the accuracy of the jitter simulation.
\item We will add a separate module enabling the use of a photon counting detector instead of a CCD, for future FUV missions. This module will further include emission from geocoronal lines, which are not relevant in the NUV, but critical in the FUV.
\item Currently, the simulator extracts the information relative to background stars from the guide star catalog\cite{lasker}, but we have already developed a replacement for that in which the simulator extracts the data from the second data release of GAIA\cite{gaia}. We will switch to this as the default as soon as the GAIA team releases conversions of GAIA output parameters into other related quantities, such as the conversion of the GAIA magnitudes into Johnson $V$-band magnitude, making the background module more complete and robust.
\end{itemize}

\section{Acknowledgements}
%-------------------------------------
A.~G.~Sreejith, L. Fossati, and M. Stellar acknowledge financial support from the Austrian Forschun\\gsf\"orderungsgesellschaft FFG project “ACUTEDIRNDL” P859718 and "CONTROL" P865968. CUTE is supported by NASA grant NNX17AI84G (PI - K. France) to the University of Colorado, Boulder. We thanks the anonymous referees for their valuable comments.

\vspace{1ex}
%\noindent Biographies and photographs of the authors are not available.

%\listoffigures
%\listoftables
\end{spacing}
\end{document}